\documentclass[onecolumn,10pt,superscriptaddress]{revtex4-2}
\pdfoutput=1
\usepackage[final]{graphicx}
\usepackage{booktabs}
\usepackage{times,bbm,amsmath,amssymb}
\usepackage{epsfig,color}
\usepackage[dvipsnames]{xcolor}
\usepackage{hyperref}
\usepackage{float}
\usepackage[caption = false]{subfig}
\usepackage{thumbpdf,enumerate}
\usepackage{booktabs}
\usepackage{sidecap}
\usepackage{pstricks}
\usepackage{multirow}
\usepackage{placeins}
\usepackage{pst-grad,bm}
\usepackage{epigraph}
\usepackage{longtable}
\usepackage{soul}
\usepackage{braket}
\usepackage{acronym}
\usepackage{comment}
\setcitestyle{round}
\usepackage[dvipsnames]{xcolor}     
\usepackage[normalem]{ulem}

\begin{document}
\title{Quantum walks of two correlated photons in a 2D synthetic lattice} 

\author{Chiara Esposito}
\affiliation{Dipartimento di Fisica, Sapienza Universit\`{a} di Roma, Piazzale Aldo Moro 5, I-00185 Roma, Italy}
\author{Mariana R. Barros}
\affiliation{Dipartimento di Fisica, Sapienza Universit\`{a} di Roma, Piazzale Aldo Moro 5, I-00185 Roma, Italy}
\affiliation{Dipartimento di Fisica "Ettore Pancini",  Universit\`{a} degli studi di Napoli Federico II, Complesso Universitario di Monte S. Angelo, via Cintia, 80126 Napoli, Italy}
\author{Andr\'es Dur\'an Hern\'andez}
\affiliation{Dipartimento di Fisica, Sapienza Universit\`{a} di Roma, Piazzale Aldo Moro 5, I-00185 Roma, Italy}
\affiliation{Université Paris-Saclay, ENS Paris-Saclay, Département de Physique, 91190 Gif-sur-Yvette, France}
\author{Gonzalo Carvacho}
\affiliation{Dipartimento di Fisica, Sapienza Universit\`{a} di Roma, Piazzale Aldo Moro 5, I-00185 Roma, Italy}
\author{Francesco Di Colandrea}
\affiliation{Dipartimento di Fisica "Ettore Pancini",  Universit\`{a} degli studi di Napoli Federico II, Complesso Universitario di Monte S. Angelo, via Cintia, 80126 Napoli, Italy}
\author{Raouf Barboza}
\affiliation{Dipartimento di Fisica "Ettore Pancini",  Universit\`{a} degli studi di Napoli Federico II, Complesso Universitario di Monte S. Angelo, via Cintia, 80126 Napoli, Italy}
\author{Filippo Cardano}
\email{filippo.cardano2@unina.it}
\affiliation{Dipartimento di Fisica "Ettore Pancini",  Universit\`{a} degli studi di Napoli Federico II, Complesso Universitario di Monte S. Angelo, via Cintia, 80126 Napoli, Italy}\author{Nicol\`{o} Spagnolo}
\affiliation{Dipartimento di Fisica, Sapienza Universit\`{a} di Roma, Piazzale Aldo Moro 5, I-00185 Roma, Italy}
\author{Lorenzo Marrucci}
\affiliation{Dipartimento di Fisica "Ettore Pancini",  Universit\`{a} degli studi di Napoli Federico II, Complesso Universitario di Monte S. Angelo, via Cintia, 80126 Napoli, Italy}
\author{Fabio Sciarrino}
\email{fabio.sciarrino@uniroma1.it}
\affiliation{Dipartimento di Fisica, Sapienza Universit\`{a} di Roma, Piazzale Aldo Moro 5, I-00185 Roma, Italy}
\affiliation{Consiglio Nazionale delle Ricerche, Istituto dei sistemi Complessi (CNR-ISC), Via dei Taurini 19, 00185 Roma, Italy}

\begin{abstract}

Quantum walks represent paradigmatic quantum evolutions, enabling powerful applications in the context of topological physics and quantum computation.  They have been implemented in diverse photonic architectures, but the realization of a two-particle dynamics on a multi-dimensional lattice has  hitherto been limited to continuous-time evolutions. To fully exploit the computational capabilities of quantum interference it is crucial to develop platforms handling multiple photons that propagate across multi-dimensional lattices. Here, we report a  discrete-time quantum walk of two correlated photons in a two-dimensional lattice, synthetically engineered by manipulating a set of optical modes carrying quantized amounts of transverse momentum. Mode-couplings are introduced via the polarization-controlled diffractive action of thin geometric-phase optical elements. The entire platform is compact, efficient, scalable, and represents a versatile tool to simulate quantum evolutions on complex lattices. We expect that it will have a strong impact on diverse fields such as quantum state engineering, topological quantum photonics, and Boson Sampling.

\end{abstract} 

\maketitle

\section{Introduction}
A quantum walk\citep{Venegas2012} (QW) is the quantum-mechanical analogue of the classical random walk, describing the evolution of a quantum particle that moves on a discrete lattice, hopping between adjacent sites. These quantum dynamics can be either continuous or discrete in time, depending on whether the couplings between neighbouring lattice positions are continuously active or can be described as sudden kicks, occurring at discrete time-steps. In the latter case, at each step the walker moves in a direction that reflects the state of  {a spin-like} internal degree of freedom, playing  the role of ``quantum coin''. The growing interest in QWs is due to their potential use in diverse quantum applications, as for instance quantum search algorithms\cite{Ambainis2004}, quantum gates for universal quantum computation\cite{Childs2009,Childs2013,Yan2019,Gong2021}, quantum state engineering\cite{Grafe2014,Pitsios2017,Giordani2019,Giordani2021}, and quantum simulations of topological and physical phenomena\cite{Kitagawa2012,kitagawa2012exp,Cardano2017,Barkhofen17,Chen18,Guo20,Nitsche2019,Geraldi19,Wang2019,Wang2019top,Erhardt2020,DErrico2020,Geraldi2021}.

Quantum walk dynamics exhibit a richer variety of phenomena when the evolution involves more than one particle. These walks indeed are characterized by multi-particle interferences \cite{Sansoni2012,Crespi2013,Carolan14,Preiss2015, Laneve2021}, having no classical analogue and incorporating an inherent source of complexity, as highlighted by their central role in computational models such as Boson Sampling\cite{AA10,Brod19review}. The dimensionality of the lattice where the QW takes place is also a crucial ingredient. As an example, quantum search algorithms based on QWs overcome their classical counterparts solely when the spatial dimension of the lattice is equal or greater than two\cite{Tulsi2008}. Furthermore, two-dimensional (2D) QWs display a richer landscape of topological features, when compared to the 1D case\cite{Kitagawa2012,Asboth2015}.

\begin{figure}[b!]
    \includegraphics[width=\textwidth]{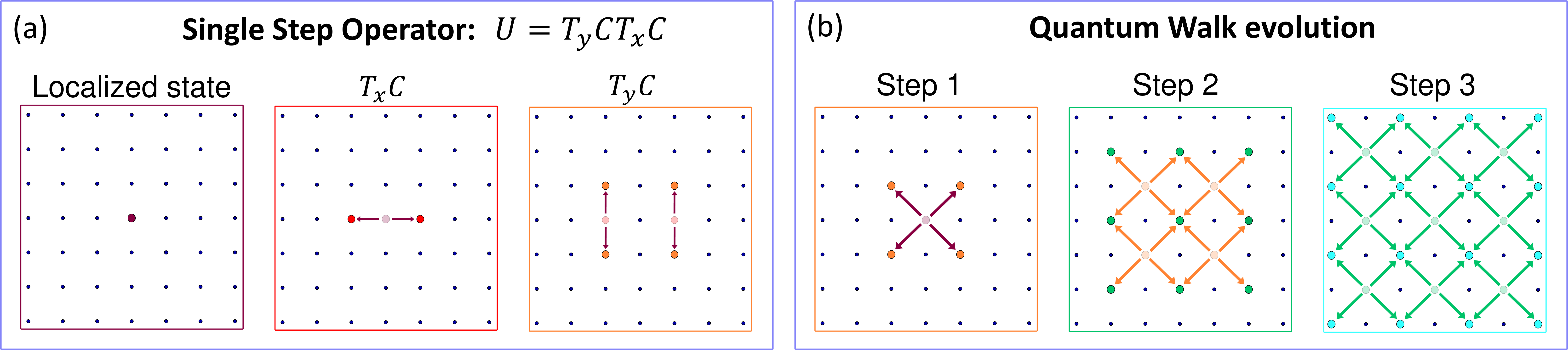}
    \caption{\textbf{Scheme of two-dimensional quantum walk.} (a) The single-step operator is performed via subsequent applications of a coin-dependent translation along the $x$ axis and one along the $y$ axis, interspersed with a coin rotation. (b) The full quantum walk evolution is then obtained via the multiple sequential application of the single-step operator $U$.}
    \label{fig:1}
\end{figure}

\begin{figure*}[t!]
    \includegraphics[width=\textwidth]{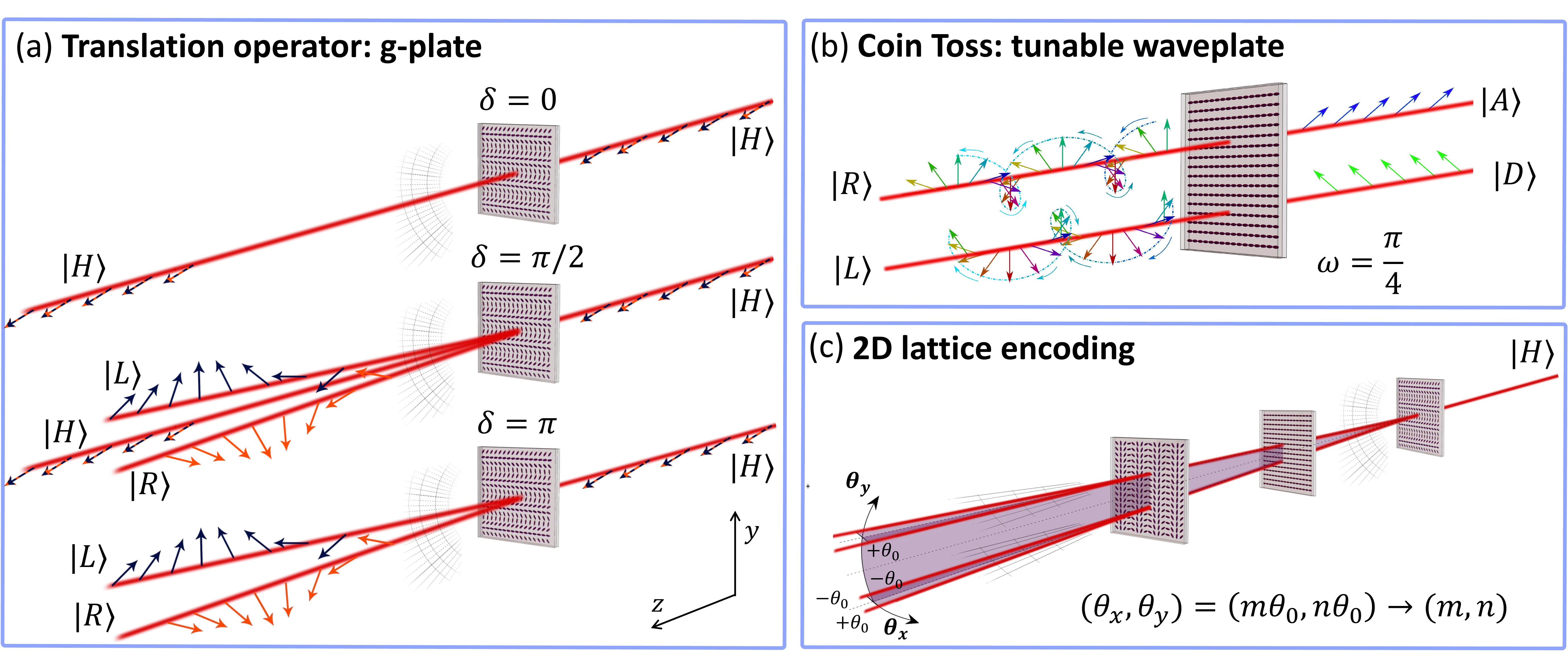}
    \caption{\textbf{Encoding of QW operators and walker degree of freedom.}  \textit{(a) Translation operator: g-plate.} Action of a g-plate oriented along $y$ for different values of $\delta$. For $\delta=0$ the g-plate acts as the identity operator and the light beam is unchanged, while for $\delta=\pi$ the g-plate performs full conversion. Finally, for all the other values of $\delta$ the g-plates convert only partially the input beam.  \textit{(b) Coin toss: tunable waveplate.} A waveplate with tunable retardation performs the coin rotation. In the balanced case (shown in the figure), the coin toss is performed by setting $\omega = \pi/4$ (quarter waveplate), implementing for instance the transformations $\vert R \rangle \rightarrow \vert A \rangle =  (\vert H \rangle - \vert V \rangle)/\sqrt{2}$ and $\vert L \rangle \rightarrow \vert D \rangle =  (\vert H \rangle + \vert V \rangle)/\sqrt{2}$. \textit{(c) 2D lattice encoding.} The walker position on the 2D lattice is encoded in transverse momentum of the beam. We consider a linearly polarized input beam, without transverse momentum component. The corresponding position on the lattice is (0,0). A g-plate, oriented along $y$ ($x$) with $\delta=\pi$, divides a horizontally-polarized input beam in two parts, which acquire two opposite transverse momentum components along $y$ ($x$), $k_{y(x)}=\pm \Delta k_\perp$ or $\theta_{y(x)} =\pm\theta_0$. Then, after coin tossing, a second g-plate is oriented along the orthogonal direction. Four different beams with different angular deviations $(\pm\theta_0,\pm\theta_0)$ are obtained at the output, mapping the lattice positions $(m,n)=(\pm1,\pm1)$, respectively. The red lines on the images indicate the propagation direction of the Gaussian beams. The beam deviations have been overemphasized for the sake of visualization. The deflected beams remain spatially overlapping while they travel along the setup, except in the final imaging stage.}
    \label{fig:concept}
\end{figure*}

Photonic platforms developed to implement QWs differ in terms of the methods to encode both walker and coin systems into optical degrees of freedom. Starting from the first experiments in linear optical interferometers composed of beamsplitters and phase shifters\cite{Broome2010}, integrated photonic technology has enabled significantly larger instances both in their continuous-time\cite{Carolan14, Carolan15, Qiang2016, Defienne2016, Tang2018, Erhardt2020, Jiao2021} and discrete-time version\cite{Crespi2013,Pitsios2017,Harris2017,Imany2020}. Other schemes rely on light polarization and orbital angular momentum degrees of freedom \cite{Cardano2015,Giordani2019}, multimode fibers\cite{Defienne2016} or fiber network loops\cite{Schreiber2010, Schreiber2012,Barkhofen17,Chen18,Weidemann2020}, where the walker position is simulated by the temporal separation between the laser pulses. Other remarkable experiments have been conducted by controlling confined waves in arrays of micro-resonators \cite{Mittal2016,Zhao2019}.  Multiparticle regimes have been already implemented in continuous-time QWs \cite{Szameit2010, Peruzzo2010, Poulios2014}, even in 2D lattices \cite{Jiao2021}. However, the demonstration of multi-photon discrete-time QWs in more than one spatial dimension has remained elusive thus far. Here we devise and experimentally validate  a compact, flexible and scalable photonic platform that achieves this goal. Specifically, we realize a three-step quantum walk dynamics with coherent light, one-photon and two-photon inputs, spanning 2, 6 and 12 modes for the first, second and third step, respectively.\\

\section{Results}
\subsection{Model and encoding}

\noindent The essential elements of a discrete time QW are captured by the single-step evolution operator $U$, as after an evolution of $t$ time-steps the system is described by a quantum state $\ket{\psi_t}=U^t\ket{\psi_0}$, where $\ket{\psi_0}$ is the input state. The operator $U$ typically includes a spin rotation $C$, acting only on the coin Hilbert space, and a spin-dependent shift. When the walker moves on a 2D square lattice [see Fig.\ \ref{fig:1}], the conditional shift embeds translation operators $T_x$ and $T_y$ along $x$ and $y$ directions , respectively (more details can be found in Supplementary Note 1).
\begin{figure*}
\includegraphics[width=\textwidth]{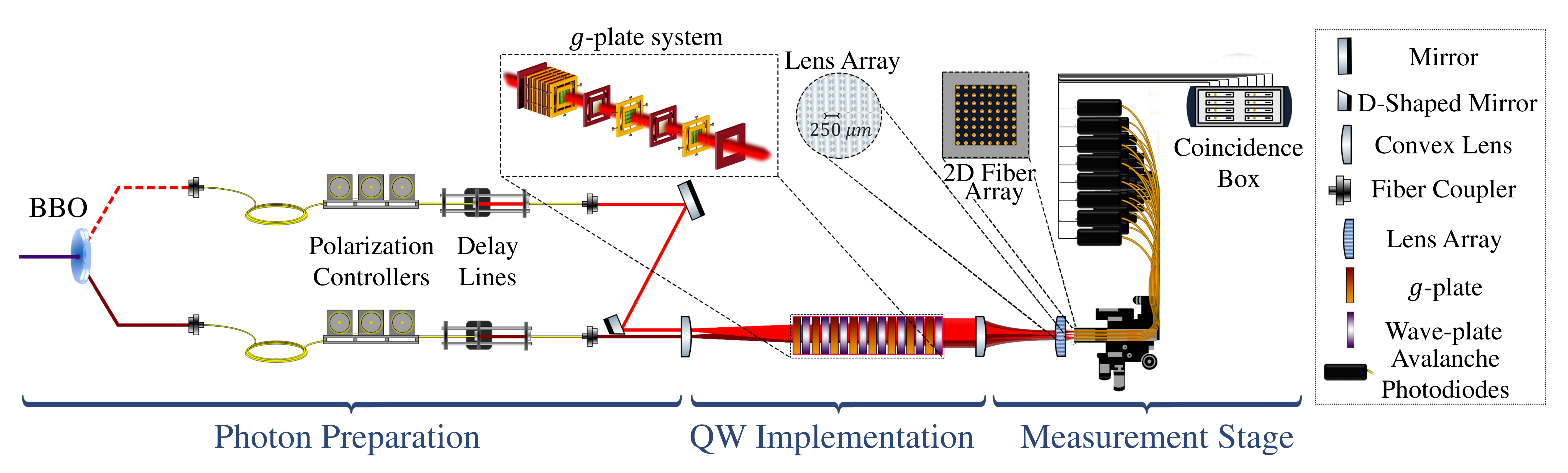}
\caption{\textbf{Experimental apparatus for the 2D-QW implementation.} \textit{Photon preparation.} Two photons are generated by a spontaneous parametric down-conversion source and independently injected into single-mode fibers. Polarization controllers are employed to change their polarization state, while delay lines enable controlling their degree of indistinguishability.  For two-photon inputs, both photons are injected in the QW implementation and they propagate along two parallel paths. A lens system enlarges their waist radius and introduces a relative angle between the optical modes. The relative inclinations of the optical modes represent different lattice positions. In this way the photons start the walk at positions $(-1,0)$ and $(1,0)$ of the lattice.In the single-particle case, one of the two photons is directly measured to act as a trigger. \textit{QW implementations.} The quantum walk is performed by using waveplates and g-plates arranged in a cascade configuration. Each g-plate is controlled independently by tuning its phase retardation via a voltage controller. \textit{Measurement stage.} A lens system at the output stage converts the different momentum values into a spatial grid. Then, for single photon acquisition, an array of micro-lenses is used to efficiently inject the output modes in a 2D square-lattice fiber array. Finally, each output fiber is plugged into an avalanche photo-diode detector connected to a coincidence electronic system. For the coherent light data acquisition, we inserted a beamsplitter between the last lens and the micro-lense array, and we positioned the CCD on the reflected path to perform image acquisition.}
    \label{fig:scheme}
\end{figure*}
Our platform builds on a recent approach to the simulation of single-particle 2D QWs using coherent laser light \cite{DErrico2020}. Here, the walker positions are provided by optical modes $\ket{m,n}$ with the following spatial profile:
\begin{equation}
\label{eq:gaussianbeamtilted}
f_{m,n}(x,y,z) = A(x,y,z) e^{i[\Delta k_\perp (m x+n y) +k_z z]},
\end{equation}
where $A(x,y,z)$ is a Gaussian envelope with large beam radius $w_0$ in the transverse $xy$ plane, $\Delta k_\perp$ represents a quantum of transverse momentum and the $z$ axis is regarded as the main propagation direction. $\Delta k_\perp$ fulfils the condition $\Delta k_\perp\ll2\pi/\lambda$, $\lambda$ being the optical wavelength. Modes in Eq.\ref{eq:gaussianbeamtilted} are essentially Gaussian beams propagating along a direction that is slightly tilted with respect to the main propagation axis [see Fig.\ \ref{fig:1}(a)]. Photons associated with mode $\ket{m,n}$ carry an average transverse momentum $\langle(k_x,k_y)\rangle=(m\Delta k_\perp,n\Delta k_\perp)$.\\
Basis states of the coin space ($\ket{\downarrow}$, $\ket{\uparrow}$) are encoded in right-handed ($\ket{R}$) and left-handed ($\ket{L}$) circularly polarized states, respectively. The conditional shift is realized via a liquid-crystal polarization grating, that is a g-plate \cite{DErrico2020}. These plates are made of a thin layer of liquid crystal, whose molecular orientation is arranged in a periodic pattern along one of the directions that are transverse to the propagation direction. Considering for instance a g-plate with a modulation along the $x$ direction, with a period $\Lambda=2\pi/\Delta k_\perp$, its action on spatial modes defined in Eq.\ \ref{eq:gaussianbeamtilted}, combined with circular polarized states, has the following expression:

\begin{equation}
\label{eq:1}
\ket{m,n,L/R
} \rightarrow \cos\frac{\delta}{2}\ket{m,n,L/R}+i \sin\frac{\delta}{2}\ket{m\pm1,n,R/L}.
\end{equation}

\noindent A similar expression holds in case the modulation is along the $y$ axis. The parameter $\delta$ is the birefringent optical retardation of the plate, that can be adjusted by applying an external voltage to the outer surfaces of the liquid crystal cell \cite{Piccirillo2010}.  Thus, in our encoding the g-plates implement the generalized shift operators $T_x(\delta)$ and $T_y(\delta)$, with the value of $\delta$ determining the fraction of the wavefunction that is shifted to neighbouring sites [see Fig.\ \ref{fig:1}(a)]. We use optical waveplates with tunable retardation $\omega$ (based on uniformly patterned liquid-crystals) to implement adjustable coin rotations $C(\omega)$ [see Fig.\ \ref{fig:concept}(b)]. Therefore, the single-step operator is an ordered sequence of g-plates and waveplates. A large variety of QWs can be implemented with this platform, by tuning the parameters $\delta$ and $\omega$, as already shown in Ref.\ \cite{DErrico2020}. In this work, we will  {focus on} a fully-balanced 2D-QW protocol described by the following one-step operator $U= T_y(\pi) C(\pi/4) T_x(\pi) C(\pi/4)$ [see Fig.\ \ref{fig:concept}(c)].
 
\subsection{Experimental setup}
\noindent The complete setup is shown in Fig.\ \ref{fig:scheme} (more details can be found in Supplementary Notes 2-4 and Supplementary Figures 1-3). A photon-pair source is employed to generate and then inject single- and two-photon inputs into the quantum walk platform. In particular, in the two-photon case, an appropriate optical system is implemented to inject the two particles in different sites of the lattice, corresponding to different optical modes (see Methods). The same apparatus can be used to inject classical laser light. The quantum walk itself is implemented via a cascade of the single-step building blocks described above. Note that, by turning on and off individual g-plates in sequence, that is by setting $\delta$ to $\pi$ or $0$, respectively, it is possible to measure the spatial distribution of the walker after each step of the protocol.  Finally, the output of the quantum walk is sent to the detection stage. In the focal plane of a lens, our modes can be spatially resolved as they form a grid of small spots. A suitable 2D fiber array and a micro-lens system are placed at this position, so that each spot matches the core of the corresponding fiber in the array  (see Methods). With a classical light input, the output can be measured via a charge-coupled device (CCD) camera.

\begin{figure*}
    \includegraphics[width=0.95\textwidth]{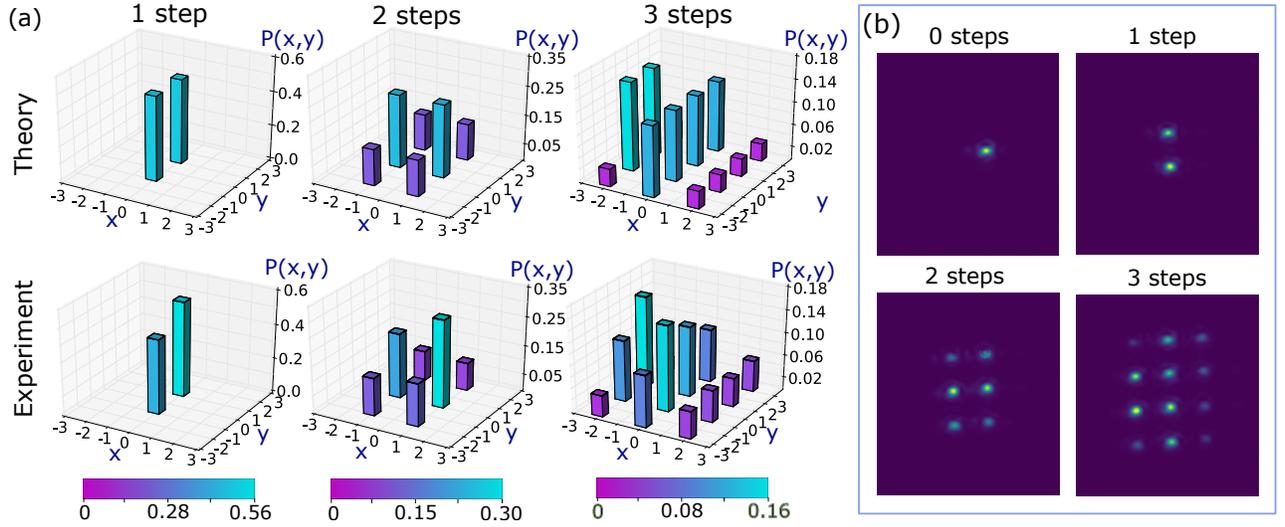}
    \caption{\textbf{Single-photon 2D quantum walk.} (a) Experimental distribution of one-particle quantum walk performed with single-photon input, in comparison with theoretical predictions after each step. Data were collected using single-photon detectors. The initial position of the walker is $(1,0)$ and the initial polarization is $\ket{D}$. Shaded regions on top of each bar correspond to the experimental error at 1 standard deviation.  {The error bars were obtained through a bootstrapping approach.} (b) Images reconstructed with a CCD camera with classical light inputs on the same site and with the same polarization.}
    \label{fig:sim1P}
\end{figure*}
\begin{figure*}
\centering
\includegraphics[width=1\textwidth]{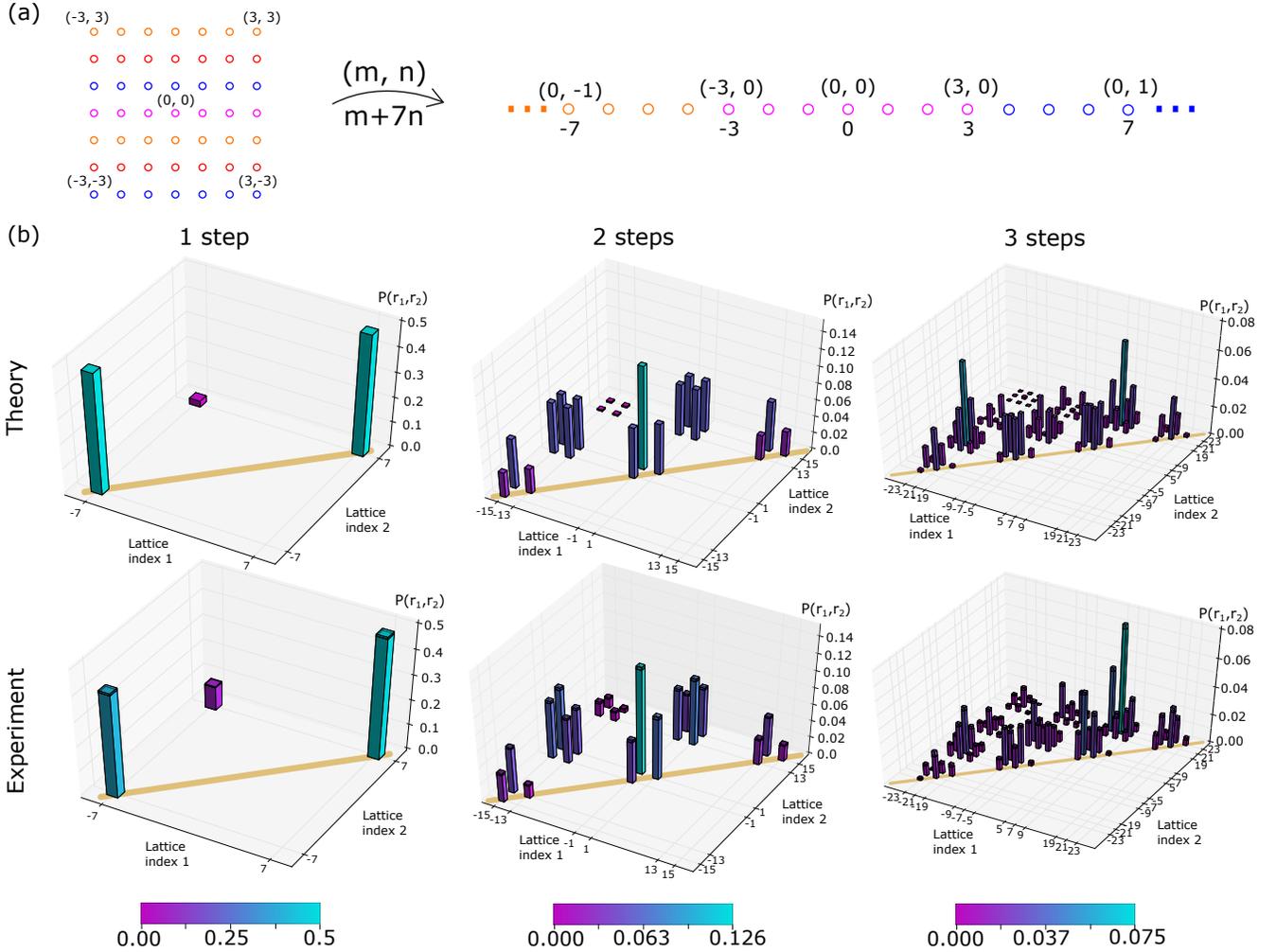}
\caption{\textbf{Experimental results for two-photon 2D-QW.} (a) In order to obtain a three-dimensional representation of the distributions, the two-dimensional lattice was linearized in the following way: $(m,n) = m + 7n$, with $m,n \in [-3,3]$. The figure graphically shows the linearization map. (b) Comparison between the theoretical predictions and the measured experimental distributions with two-photon inputs. Shaded regions on top of each bar correspond to the experimental errors. The error bars were obtained through a bootstrapping approach. The bunching probabilities are highlighted by the yellow line.}
\label{fig:sim2P}
\end{figure*}

\noindent We have employed our platform to implement a balanced quantum walk $U$ with single- and two-photon inputs, up to three time-steps. 

\subsection{Single-photon Quantum Walk}
We performed a single-particle experiment by injecting a single photon in position (1,0) with polarization $\ket{D}= (\ket{H}+\ket{V})/\sqrt{2}$, and then measuring the output distributions after each step (obtained by sequentially switching on the g-plates). The single-photon experimental distributions are shown in Fig.\ 4  (related analysis for coherent light input is reported in Supplementary Note 5). The slight differences between theoretical and experimental output distributions are due to experimental imperfections such as inaccuracies of g-plates tuning values and of their relative horizontal alignment. The agreement between experimental data and the expected distributions is quantified by the similarity, defined as:
\begin{equation}
    \mathcal{S}_{1\mathrm{p}}^{(t)}=\left(\sum_{\mathbf{r}}\sqrt{P^{(t)}({\mathbf{r}}) \tilde{P}^{(t)}({\mathbf{r}}})\right)^2,
\end{equation}
where $P^{(t)}({\mathbf{r}})$ and $\tilde{P}^{(t)}({\mathbf{r}})$ are the theoretical and experimental distributions of the quantum walk at the $t$-th step, respectively, while $\mathbf{r}$ is the particle position on the lattice. The similarity value of the last step $\mathcal{S}_{1\mathrm{p}}^{(3)}=0.9773 \pm 0.0002$ shows a high agreement with the expected distribution. Similar results were observed when injecting classical light to the quantum walk platform (see Table \ref{tab:Results}).

\subsection{Two-photon Quantum Walk}

\noindent  Next, we realized a two-photon, 2D quantum walk by injecting two photons with polarizations $\ket{A}= (\vert H \rangle - \vert V \rangle)/\sqrt{2}$ and $\ket{D}$ in $(-1,0)$ and $(1,0)$ lattice positions, respectively. It is worth noticing that at the input of the quantum walk the two-photon state is separable.
Temporal synchronization between the particles is ensured in advance by performing a direct Hong-Ou-Mandel (HOM) measurement at the output of the quantum walk, which provides a measured visibility $v=0.95 \pm 0.02$. Further details on the HOM measurement are reported in Supplementary Note 6 and Supplementary Figure 4. For the multi-photon case, the obtained theoretical and experimental distributions depicted in Fig.\ \ref{fig:sim2P} show a high quantitative agreement. This is confirmed by their similarities defined as:
\begin{equation}
    \mathcal{S}_{2\mathrm{p}}^{(t)}=\left(\sum_{\mathbf{r}_{1},\mathbf{r}_{2}}\sqrt{P^{(t)}({\mathbf{r}_{1},\mathbf{r}_{2}}) \tilde{P}^{(t)}({\mathbf{r}_{1},\mathbf{r}_{2}}})\right)^2.
\end{equation}

where $P^{(t)}({\mathbf{r}_{1},\mathbf{r}_{2}})$ and $\tilde{P}^{(t)}({\mathbf{r}_{1},\mathbf{r}_{2}})$ are the theoretical and experimental distributions of the quantum walk at the $t$-th step, respectively, while $\mathbf{r}_{1}$ and $\mathbf{r}_{2}$ are the position on the lattice of the two particles. The theoretical distribution was computed by considering an initial state described by the density matrix $\rho_0=c_0\rho_{\text{ind}}+(1-c_0)\rho_{\text{dis}}$, where $\rho_{\text{ind}}$ indicates the density matrix of two completely indistinguishable photons and $\rho_{\text{dis}}$ is the density matrix describing two distinguishable particles. $c_0$, that corresponds to the measured visibility of the HOM test, is equal to 0.95. For the experimental distribution at the third step we obtain a similarity value  $\mathcal{S}_{2\mathrm{p}}^{(3)}=0.914 \pm 0.002$
(related results with a distinguishable two-photon input are reported in Supplementary Note 7 and Supplementary Figure 5).

\subsection{Non-classical correlation witness}

\begin{figure*}
\centering
\includegraphics[width=\textwidth]{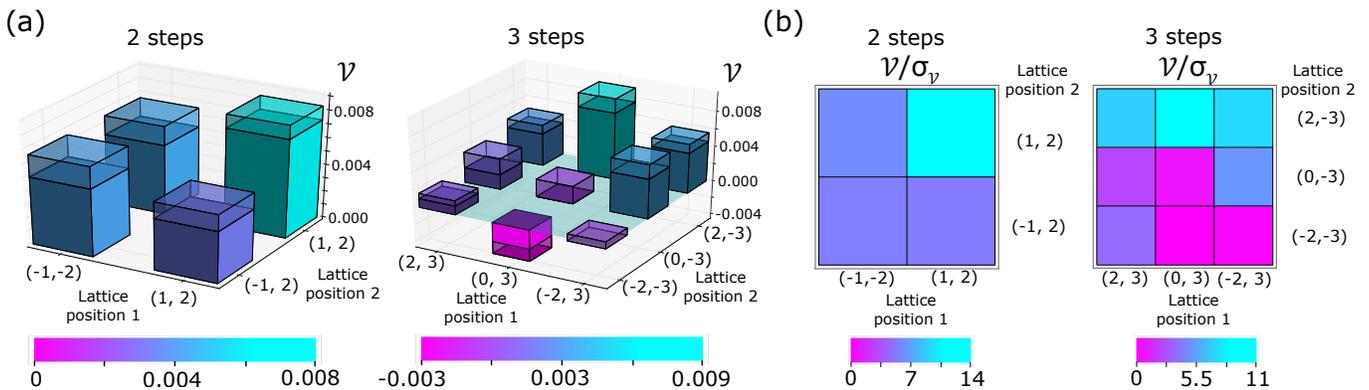}
\caption{\textbf{Violation for second and third step.} (a) Plots of $\mathcal{V}(\mathbf{r}_{1},\mathbf{r}_{2})$ with their error $\sigma_{V}(\mathbf{r}_{1},\mathbf{r}_{2})$ for the second and third step, for the lattice sites that satisfy the conditions explained in Supplementary Note 7. (b) Plots of $\mathcal{V}(\mathbf{r}_{1},\mathbf{r}_{2})/\sigma_{\mathcal{V}}(\mathbf{r}_{1},\mathbf{r}_{2})$, for those lattice sites where such quantity is positive, thus violating Eq.\ \ref{eqn:viol} for the second and the third step. }
   \label{fig:Vio}
\end{figure*}

\noindent The presence of non-classical correlations in the output distributions is witnessed by applying the non-classicality test of Refs.\ \cite{Bromberg2009,Peruzzo2010}, given by:
\begin{equation}
    \mathcal{V}({\mathbf{m}_{1},\mathbf{m}_{2}}) = \frac{2}{3} \sqrt{\Gamma_{\mathbf{m}_{1},\mathbf{m}_{1}}^{(\rm{cl})}\Gamma_{\mathbf{m}_{2},\mathbf{m}_{2}}^{(\rm{cl})}}-\Gamma_{\mathbf{m}_{1},\mathbf{m}_{2}}^{(\rm{cl})} < 0.
    \label{eqn:viol}
\end{equation}
where $\Gamma_{\mathbf{m}_1,\mathbf{m}_2}^{(\rm{cl})}$ is the classical probability that light exits from the $\mathbf{m}_1$ and $\mathbf{m}_2$ output ports of an interferometer. How this formula adapts to our specific case is shown in detail in the Supplementary Note 7. The obtained maximum violation of the inequality are of 96, 14 and 11 standard deviations, respectively for 1, 2 and 3 QW steps, thus unambiguously proving the quantum behavior of the reported two-photon 2D quantum walk.  The  indistinguishability between injected photons gives rise to quantum interferences, yielding in turn probability distributions that cannot be reproduced by classical states of light. In Fig.\ \ref{fig:Vio}, we report the complete plots of $\mathcal{V}$ over the standard deviation for the second and third step, respectively. All the results of one-particle and two-particle QWs are summarized in Table \ref{tab:Results}. In the last column, we also reported the maximum value of the violation $\mathcal V$ and its error, for each step.
These results highlight that the proposed platform has the potential to be employed for significantly larger instances, with high degree of control on the implemented protocol.

\begin{table}
   \centering
    \begin{tabular}{cccccc}
    \hline
  & $\mathcal{S}_{\mathrm{cl}}$&  $\mathcal{S}_{1\mathrm{p}}$ & $\mathcal{S}_{2\mathrm{p}}$  & $\mathcal{V}$ & $\mathcal{V}/\sigma_\mathcal{V}$\\
    \hline
    \textbf{1 step} &\, $\;\sim0.999\;$ &\,$\;0.9964(1)\;$   & $\; {0.9756(6)}\;$ & $\; {0.204(2)}\;$ & $\; {96}\;$ \\
    \textbf{2 steps} &\,$\;\sim 0.996\;$ &\,$\;0.9929(2)\;$  & $\; {0.9743(7)}\;$ & $\; {0.0077(5)}\;$& $\; {14}\;$ \\
    \textbf{3 steps} & $\;\,\sim0.988\;$ &\,$\;0.9773(2)\;$  & $\; {0.914(2)}\;$  & $\; {0.0084(7)}\;$& $\; {11}\;$  \\
    \hline
    \end{tabular}
    \caption{\textbf{Similarities of distributions related to classical, one particle and two particle regime for all the steps.} Experimental results for the 2D quantum walk up to the 3rd step when using classical light ($\mathcal{S}_{\mathrm{cl}}$) one photon ($\mathcal{S}_{1\mathrm{p}}$) and two photons ($\mathcal{S}_{2\mathrm{p}}$). In the last columns we report the maximum value of $\mathcal{V}$ with the  associated errors $\sigma_\mathcal{V}$, and the corresponding values of $\mathcal{V}/\sigma_\mathcal{V}$ for each measured configuration.}
     \label{tab:Results}
\end{table}

\section{Discussion}
\noindent We have presented and realized a platform for the implementation of two-dimensional, multiphoton discrete-time quantum walks, demonstrating experimentally single- and two-photon operations on a 2D  {squared} lattice. The presented platform is compact, flexible and enables the implementation of a large variety of different topological quantum walks \cite{DErrico2020}. Hence, it can represent a powerful tool for the investigation of rich dynamics that are experimentally unexplored. {Recent works reported 2D continuous-time quantum walks of correlated photons, relying on arrays of coupled waveguides \cite{Erhardt2020,Jiao2021, Qiang2021}} {, and important results have been also achieved by means of superconductive quantum processors \cite{Gong2021}}. We stress that our system is based on a very different approach, exploiting a synthetic 2D lattice made of internal modes of a single optical beam, as opposed to real-space neighbouring lattice sites, and implementing discrete-time evolutions that can be actively controlled and easily reconfigured. We believe that these different approaches may have complementary advantages.  {In our platform, several quantum walk protocols can be dynamically realized. This can be achieved by tuning the retardation of each plate in the range $[0,\pi]$, by changing their orientation and their position in the plane transverse to the photons main propagation direction. By controlling these parameters,  diverse single particle QWs mimicking periodically driven Chern insulators have been reported \cite{DErrico2020}. Here we implement a different split-step quantum walk, that is proved to realize the Grover search algorithm in the high step-number limit \cite{Difranco2011}. The number of steps that can be currently realized is essentially limited by the optical losses, which  {are mainly due to photon reflections at} each plate. However, these can be significantly reduced (from $\sim 15\%$ to at least $\sim 5\%$) by adding a standard anti-reflection coating on the plate outer surfaces.} Several applications can be foreseen, including quantum state engineering \cite{Grafe2014,Pitsios2017,Giordani2019} or quantum algorithms based on the quantum-walk paradigm \cite{Ambainis2004,Childs2009}. Furthermore, given the possibility to exploit multi-photon inputs and to control the performed transformation, this approach can also represent a promising platform for the implementation of Boson Sampling and Gaussian Boson Sampling experiments in large optical lattices \cite{Brod19review}.

\section{Methods}

\subsection{Photons preparation}

\noindent The photon pairs used in our experiment are generated by a parametric down-conversion source, composed of a nonlinear beta barium borate crystal (BBO) pumped by a pulsed laser with $\lambda=392.5$ nm. The generated photons, $\lambda=785$ nm, are then injected in two identical single mode fibers for spatial mode selection. On each fiber, an independent polarization controller allows choosing the polarization of each input photon. Then, delay lines are used to temporally synchronize the optical paths through a Hong-Ou-Mandel interference measurement.

Once the photons are temporally indistinguishable, they are injected in the QW platform. In order to precisely control the distance between the injected photons and match the coupling conditions of the fiber array, a half-mirror on one of the two paths is used. By translating this mirror, the distance between the paths can be modified from 3 mm to 8 mm, while its tilting can change the relative orientation between the two photons. Finally, an appropriate lens system superposes the two paths and introduces a small prescribed angle between them  (see Supplementary Note 4). This relative angle corresponds to into a difference in the photon transverse momentum.

\subsection{Measurement stage}

\noindent At the output of the quantum walk structure, a three-lens system is used to decrease the relative distance between adjacent beams and reduce the corresponding beam waists. Then, a micro-lens array, composed of a hundred micro-lenses with a short effective focal distance (approximately 5 mm), separated by 250 $\mathrm{\mu}$m, is used to inject the photons into a square-lattice multimode fiber array, reaching an individual waist of 15 $\mathrm{\mu}$m without changing the corresponding distances. The adoption of the micro-lens array enables an improvement in the coupling efficiency from $0.1$ to $0.75$. Overall, the fiber array sets a $8\times8$ spatial grid, where the output fibers are separated by a 250 $\mathrm{\mu}$m pitch. Each of these multi-mode fibers is connected to a single-photon avalanche photo-diode. The output signal of each detector is directed to a coincidence apparatus, able to record single-photon counts and two-fold coincidences.

\section{Data availability} 
\noindent The data that support the findings of this study are available from the corresponding authors upon reasonable request.
\section{Code availability} 
\noindent The codes for data processing are available from the corresponding authors upon reasonable request.

\section{Acknowledgments} 

\noindent The authors acknowledge D. Poderini for software assistance and G. Amico for valuable technical support. This work was supported by the European Union Horizon 2020 program, within the European Research Council (ERC) Grant No. 694683, PHOSPhOR; by project PRIN 2017 ``Taming complexity via QUantum Strategies a Hybrid Integrated Photonic approach" (QUSHIP) Id. 2017SRNBRK. 

\section{Author contributions}

\noindent C.E., G.C., N.S., F.C., L.M., F.S. conceived the project. C.E., M.R.B., A.D.H., G.C., N.S. performed the experiment. C.E., M.R.B., G.C., F.C., N.S., L.M., F.S. performed the data analysis. F.D.C., R.B., F.C., L.M. developed the g-plate devices. 
All authors discussed the results and participated in preparing the manuscript. 

\section{Competing interests}
\noindent The authors declare no competing interest.

\end{document}

\title{Supplementary Information: Quantum walks of two correlated photons in a 2D synthetic lattice}

\author{Chiara Esposito}
\affiliation{Dipartimento di Fisica, Sapienza Universit\`{a} di Roma, Piazzale Aldo Moro 5, I-00185 Roma, Italy}
\author{Mariana R. Barros}
\affiliation{Dipartimento di Fisica, Sapienza Universit\`{a} di Roma, Piazzale Aldo Moro 5, I-00185 Roma, Italy}
\affiliation{Dipartimento di Fisica "Ettore Pancini",  Universit\`{a} degli studi di Napoli Federico II, Complesso Universitario di Monte S. Angelo, via Cintia, 80126 Napoli, Italy}
\author{Andr\'es Dur\'an Hern\'andez}
\affiliation{Dipartimento di Fisica, Sapienza Universit\`{a} di Roma, Piazzale Aldo Moro 5, I-00185 Roma, Italy}
\affiliation{Université Paris-Saclay, ENS Paris-Saclay, Département de Physique, 91190 Gif-sur-Yvette, France}
\author{Gonzalo Carvacho}
\affiliation{Dipartimento di Fisica, Sapienza Universit\`{a} di Roma, Piazzale Aldo Moro 5, I-00185 Roma, Italy}
\author{Francesco Di Colandrea}
\affiliation{Dipartimento di Fisica "Ettore Pancini",  Universit\`{a} degli studi di Napoli Federico II, Complesso Universitario di Monte S. Angelo, via Cintia, 80126 Napoli, Italy}
\author{Raouf Barboza}
\affiliation{Dipartimento di Fisica "Ettore Pancini",  Universit\`{a} degli studi di Napoli Federico II, Complesso Universitario di Monte S. Angelo, via Cintia, 80126 Napoli, Italy}
\author{Filippo Cardano}
\email{filippo.cardano2@unina.it}
\affiliation{Dipartimento di Fisica "Ettore Pancini",  Universit\`{a} degli studi di Napoli Federico II, Complesso Universitario di Monte S. Angelo, via Cintia, 80126 Napoli, Italy}\author{Nicol\`{o} Spagnolo}
\affiliation{Dipartimento di Fisica, Sapienza Universit\`{a} di Roma, Piazzale Aldo Moro 5, I-00185 Roma, Italy}
\author{Lorenzo Marrucci}
\affiliation{Dipartimento di Fisica "Ettore Pancini",  Universit\`{a} degli studi di Napoli Federico II, Complesso Universitario di Monte S. Angelo, via Cintia, 80126 Napoli, Italy}
\author{Fabio Sciarrino}
\email{fabio.sciarrino@uniroma1.it}
\affiliation{Dipartimento di Fisica, Sapienza Universit\`{a} di Roma, Piazzale Aldo Moro 5, I-00185 Roma, Italy}
\affiliation{Consiglio Nazionale delle Ricerche, Istituto dei sistemi Complessi (CNR-ISC), Via dei Taurini 19, 00185 Roma, Italy}

\maketitle

%%%%%%%%%%%%%%%%%%%%%%
%\mariana{Parts in blue: related to the 1 particle regime. To be easily identified in the text.}\\
%\marianaa{Parts in purple: Mariana's comments}\\
%\gonzalo{Parts in red: Gonzalo's comments}
\section{Theoretical background}
\label{Theory}
\noindent The realization of a quantum walk on a 2D square lattice is performed by the repetitive action of an unitary single-step operator $U$ on a quantum system. Tracing a parallel with the classical random walk, the quantum system plays the role of the walker while its internal spin-like degree of freedom represents the coin \cite{Aharonov1993}. Therefore, the walker can be described by a quantum state $\ket{\psi}$ in the bipartite Hilbert space $\mathcal{H}=\mathcal{H}_p\otimes \mathcal{H}_c$. The two dimensional internal degree of freedom space, $\mathcal{H}_c$, is spanned by the states $\{\ket{\uparrow},\ket{\downarrow}\}$, and the position space $\mathcal{H}_p$ by $\{\ket{m,n}\}$, where $(m,n)$ label the square lattice coordinates. Thus, the discrete-time quantum walk evolution at the $t$-th step can be expressed as:
\begin{equation}
    \ket{\psi_t}=U^t\ket{\psi}.
\end{equation}
The single-step operator $U$  encompasses the spin rotation $C(\omega)$, which only acts on the coin space, and the spin-dependent translation operators along $x$ and $y$ directions, $T_d$, with $d=x,y$. 
The general unbalanced spin rotation is given by :
\begin{equation}
    C(\omega)= \mathbb{I}_p \otimes
    \begin{pmatrix}
    \cos (\omega) & \imath \sin (\omega) \\
     \imath \sin (\omega) & \cos (\omega) \\
    \end{pmatrix},
\end{equation}
where parameter $\omega$ sets the balance of the coin rotation. In a balanced coin rotation, there is an equal probability to move in each direction, condition achieved if $\omega=\pi/4$.

\noindent The translation operator  can be expressed as:
\begin{equation}
T_d(\delta)= \cos{\frac{\delta}{2}} \mathbb{I} +     \imath \sin{\frac{\delta}{2}} S_d,
\end{equation}
where $\delta$ establishes the probability of the walker to remain in the same position at each step, while the operator $S_d$ is the shift operator along the $d=x,y$ direction. The $S_x$ operator does not act on $y$ coordinates and it is given by:
\begin{equation}
S_x= \sum_{m,n} \ket{m-1,n,\uparrow}\bra{m,n,\downarrow}+ \ket{m+1,n,\downarrow}\bra{m,n,\uparrow}.
\end{equation}
Similarly, the operator $S_y$  does not act on $x$ coordinates and is defined as $S_y= \sum_{m,n} \ket{m,n-1,\uparrow}\bra{m,n,\downarrow}+ \ket{m,n+1,\downarrow}\bra{m,n,\uparrow}$.
The ordering of these operators in the single-step operator $U$ and the chosen values for $\delta$ and $\omega$ allows to define different QWs protocols. 

\noindent Here, we mainly focused on the 2D-QW protocol $U= T_y(\delta) C(\omega) T_x(\delta) C(\omega)$, setting $\delta = \pi$ and $\omega=\pi/4$. By making this choice, we selected a protocol in which the walker must change its position at each step,  with a balanced coin rotation.

\noindent In the particular case of one-particle regime, we considered a family of localized initial states $\ket{\psi_0}$ described by: 
\begin{equation}
    \ket{\psi_0} =\cos{\frac{\alpha}{2}}\ket{1,0,\uparrow}+\imath \sin{\frac{\alpha}{2}}\ket{1,0,\downarrow}.
\end{equation}
From the evolved state, $\ket{\psi_t}$, we obtained the probability distribution as:
\begin{equation}
P^{(t)}(\mathbf{r})=\sum_{\sigma=\uparrow,\downarrow}|\braket{\mathbf{r},\sigma|\psi_t}|^2,
\label{eqn:1inst}
\end{equation}
where the vector $\mathbf{r}$ indicates the positions $(m,n)$ in the lattice.

%From the evolved state, $\ket{\psi_t}$, we obtained the probability distribution as:
%\begin{equation}
%P^{(t)}(\vec{r})=\sum_{\sigma=\uparrow,\downarrow}|\braket{\vec{r},\sigma|\psi_t}|^2,
%\label{eqn:1inst}
%\end{equation}
%where the vector $\vec{r}$ indicates the positions $(m,n)$ in the lattice.

\noindent In the two-particle regime, partial photon indistinguishability has to be taken into account to properly model the input state and calculate theoretical predictions. For two-photon inputs, this effect can be modeled with a density matrix defined as:
\begin{equation}
    \rho_0=c_0\rho_{\text{ind}}+(1-c_0)\rho_{\text{dis}},
\end{equation}
where $\rho_{\text{ind}}$ indicates the density matrix of two completely indistinguishable photons and $\rho_{\text{dis}}$ is the density matrix describing  two distinguishable particles. The $c_0$ parameter sets the degree of indistinguishability between the two photons, which is defined in the range $[0,1]$. At each step, we obtain a probability distribution on the lattice position of the two walkers as:
\begin{equation}
P^{(t)}(\mathbf{r}_{1},\mathbf{r}_{2}) = \sum_{\sigma_1,\sigma_2=\uparrow,\downarrow}\text{Tr}(\rho_t\ket{\mathbf{r}_{1},\sigma_1,\mathbf{r}_{2},\sigma_2}\bra{\mathbf{r}_{1},\sigma_1,\mathbf{r}_{2},\sigma_2}).
\label{eqn:2inst}
\end{equation}
%\begin{equation}
%P^{(t)}(\vec{r}_{1},\vec{r}_{2}) = \sum_{\sigma_1,\sigma_2=\uparrow,\downarrow}\text{Tr}(\rho_t\ket{\vec{r}_{1},\sigma_1,\vec{r}_{2},\sigma_2}\bra{\vec{r}_{1},\sigma_1,\vec{r}_{2},\sigma_2}).
%\label{eqn:2inst}
%\end{equation}

\noindent At each step, the elaboration of the theoretical prediction for the distributions shown in the main text was performed by using Supplementary Equations \eqref{eqn:1inst} and \eqref{eqn:2inst}.

\section{g-plate Technology}

\noindent The key elements of our quantum walk are the  g-plates \cite{DErrico2020}, which are electrically controlled liquid-crystal waveplates for which the optic axis orientation is varying linearly along one of the in-plane directions. By using the same principle of the \textit{q-plates} \cite{Slussarenko2011}, an optimally tuned g-plate splits a light beam into two orthogonally polarized beams, as depicted in Fig.2 of the main text. The relative angle between the two split beams is determined by the diffraction-grating spacing of the g-plate, and it is approximately equal to $\alpha = 2\lambda/\Lambda $, where $\lambda$ is the wavelength of the input beam and $\Lambda$ the characteristic parameter of the device, defined during the fabrication process.  Thus, such device adds a fixed transverse momentum component ($\Delta k_\perp= 2\pi/\Lambda $) to an input beam, and the sign of the ``kick'' depends on its polarization state. In particular, if the beam is right-handed circularly polarized ($\ket{R}$), it acquires a negative component of the transverse momentum. On the other hand, a left-handed circularly polarized beam ($\ket{L}$) acquires a positive fixed component of the transverse momentum. The action of an $x$-oriented g-plate in the polarization space is given by the following matrix, written in the circular basis ($\ket{R}, \ket{L}$):
\begin{equation}
G_p=\begin{pmatrix}
\cos (\delta/2) & \imath \sin (\delta/2) \,e^{\imath \Delta k_\perp x}\\
\imath \sin (\delta/2) \,e^{-\imath \Delta k_\perp x} & \cos (\delta/2)
\end{pmatrix},
\end{equation}
where $\delta$ is the phase delay introduced by the g-plate, that defines each momentum ``kick" probability, and it is tunable by an external  programmable voltage source. Finally, the accumulated momentum can be visualized in position space by using a lens system that implements an optical Fourier transformation.

\noindent An overall amount of 12 electrically-tuned liquid crystal devices were exploited to perform a two-photon QW up to the third step, encompassing waveplates and g-plates arranged in a cascade configuration. In Supplementary Figure \ref{fig:gplatesystem} (a) a schematic representation of the exploited apparatus is shown, while the image of a g-plate is reported in Supplementary Figure \ref{fig:gplatesystem} (b).

\begin{figure}[ht!]
    \includegraphics[width=\textwidth]{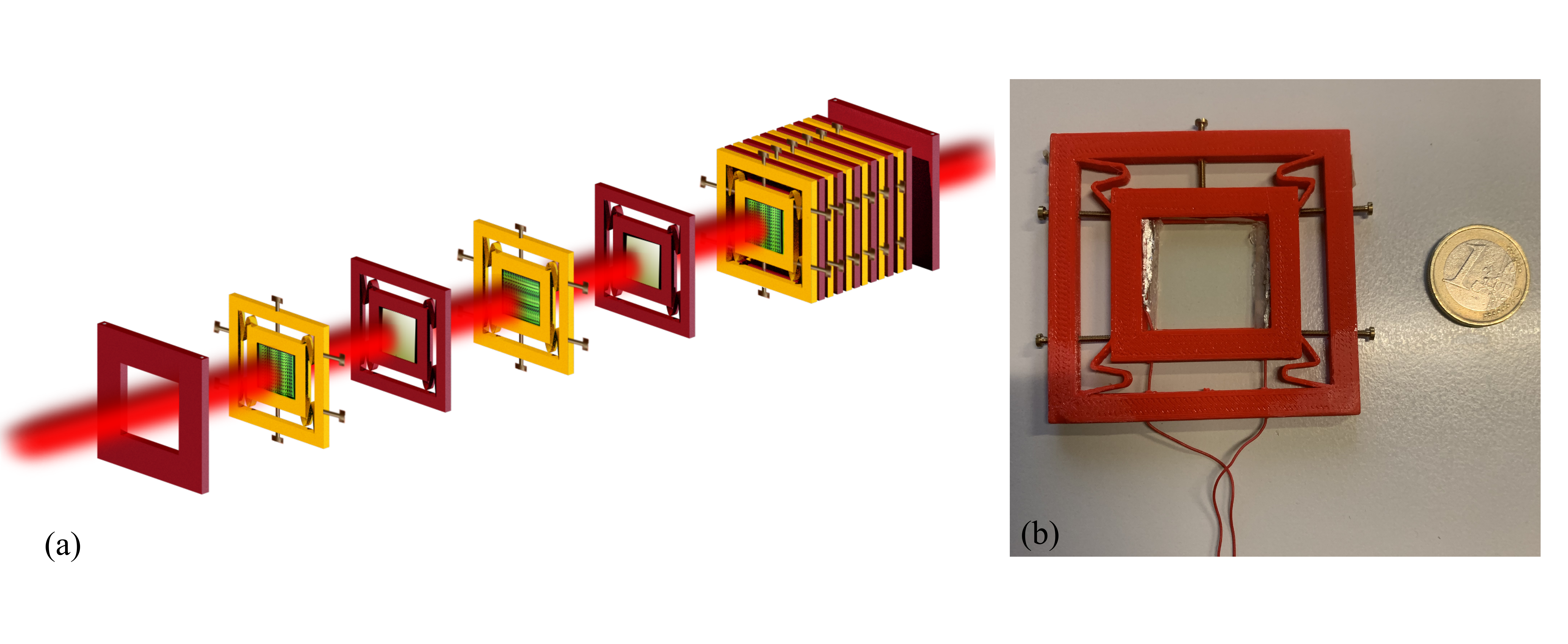}
\caption{\textbf{g-plate system.} (a) Schematic view of the quantum walk apparatus used in the experiment.  (b) Image of a single g-plate in its mount. The screws on the sides of the red support are exploited to properly align the device orientation with respect to the other g-plates.}
   \label{fig:gplatesystem}
\end{figure}

%%%%%%%%%%%%%%%%%%%%%%%%%%%%%%%%%%%%%%%%%%%%%%%%%%%%%%%%%%%%%%
\section{Micro-lens array}
\label{microlenses}
\noindent A micro-lens array, model $MLA-S250-f20$ of RPC Photonics, was used to efficiently inject the photons into the 2D fiber array employed at the detection stage. The micro-lens array is made of polymer-on-glass material, has a size of $250 \times 250$ mm$^2$, index of refraction $1.56$ ($633$ nm), and it is composed of a hundred of micro-lenses with a short effective focal length (around $5$ mm), separated by $250$ $\mu$m (see Supplementary Figure \ref{fig:microlenses}). In this way, we were able to reach an individual waist of $15$ $\mathrm{\mu}$m without changing the corresponding distances in the experiment. Consequently, the adoption of such device enabled a significant increase in the coupling efficiency into the final array of optical fibers used for detection from $0.1$ to $0.75$.

\begin{figure}[ht!]
    \centering
    \includegraphics[width=0.15\textwidth]{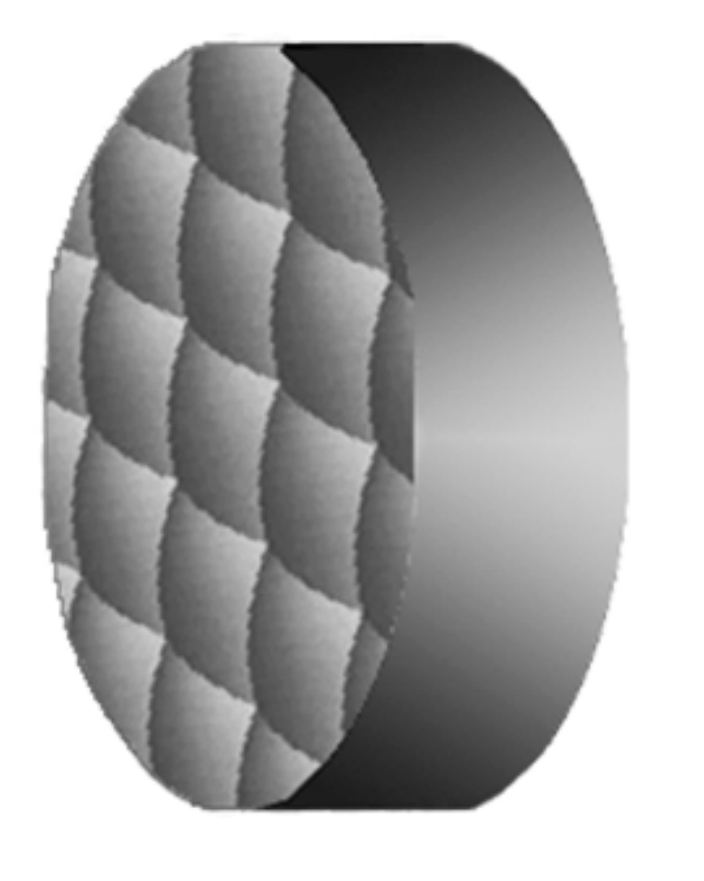}
   \caption{ \textbf{Micro-lens array.} Schematic representation of the micro-lens array used to increase the coupling efficiency of the output mode in the fiber array.}
  \label{fig:microlenses}
\end{figure}

%%%%%%%%%%%%%%%%%%%%%%%%%%%%%%%%%%%%%%%%%%%%%%%%%%%%%%%%%%%%%%
\section{Experimental implementation}

\noindent The photons used in the experiment were produced by a parametric source, consisting of a beta-barium-borate (BBO) crystal, that generates a photon pair ($\lambda=785$ nm) by non-collinear spontaneous parametric down conversion (SPDC) when it is pumped by a pulsed laser with $\lambda=392.5$ nm. Once the photons were generated by the SPDC source, they were injected into two identical single-mode fibers. Mechanical polarization controllers were employed to arbitrarily and independently control the polarization of each input photon. 

\noindent In the case of the single-particle 2D-QW, one photon was directly sent thorough a single mode fiber to an Avalanche Photodiode (APD), to be used as trigger. The other photon was sent to the platform at the fiber launch ``1'', shown in Supplementary Figure \ref{fig:sketch}. It performed the 2D-QW by following the same path as for the 2D-QW with two-photons, presented below. On the other hand, for the case of the two-photon 2D-QW, both particles were injected in the platform. In this case, temporal synchronization of the photons was performed via independent delay lines through an Hong-Ou-Mandel experiment (see Supplementary Note \ref{HOM}). 
The single-mode fibers were then connected to the fiber launchers ``1" and ``2", depicted in Supplementary Figure \ref{fig:sketch}. A mirror, mounted on a translation stage, and a half-mirror were used to finely adjust the parallelism between the two beams and to change their relative distance $d_{0}$, that can vary from $3$ mm to $8$ mm. In the sequence, a telescope, composed of two lenses with focal distances $f_1 = 300$ mm and $f_2 = 30$ mm respectively, was used to reduce the beam waist by a factor 10. The relation between the waist and distance can be expressed as following:
\begin{equation}
W_1=\frac{f_2}{f_1}W_0; \quad d_1= \frac{f_2}{f_1}d_0.
\end{equation}
Subsequently, we used a lens with focal distance $f_{3}=2000$ mm to enlarge each waist without changing the separation length. At the output of $f_3$, each beam had a waist $W_{2}$, respecting the following relations:
\begin{equation}
    W_{2} = \frac{W_1}{\Big[ 1 + (z_{1}/f_{3})^{2}\Big]^{1/2}}\,; \quad
    z_{1} = \frac{\pi W_{1}^{2}}{\lambda},
\end{equation}
where $z_1$ is the Rayleigh length of the beam after the telescope. When $z_1\ll  f_3$, this can be accurately approximated as: 
\begin{equation}
    W_2 \simeq \frac{\lambda f_3}{\pi W_1}\,\simeq \frac{\lambda}{\pi W_0}\,\frac{f_1 f_3}{f_2}.
\end{equation}
Therefore, the two beams were highly superposed and collimated. By using the ray transfer-matrix formalism, we were able to compute the relative angle $\theta_{2}$ between the two beams at the output of $f_3$, assuming an effective parallelism at the output of $f_2$:
\begin{equation}
\theta_{2} =  \frac{d_{1}}{f_3}=d_{0} \frac{f_2}{f_1f_3}.
\end{equation}
Hence, by changing the value of $d_0$, we linearly controlled the input angle between the two beams, $\theta_2$. The angular lattice unity is $\Delta\,\theta = \lambda/\Lambda$, where $\Lambda$ is the g-plate characteristic parameter defined above. For $d_0 = 3.14$ cm, the two photons are prepared at $(0,0)$ and $(1,0)$ modes of the quantum walk ($\theta_2=\lambda/\Lambda=1.57 \times 10^{-4}$ rad). Thus, increasing by a factor 2 the value of $d_0$, $\theta_2$ is also magnified by the same factor. In the latter case, the two photons were prepared at $(-1,0)$ and $(1,0)$ input lattice positions.

\begin{figure}[t!]
    \centering
    \includegraphics[width=0.99\textwidth]{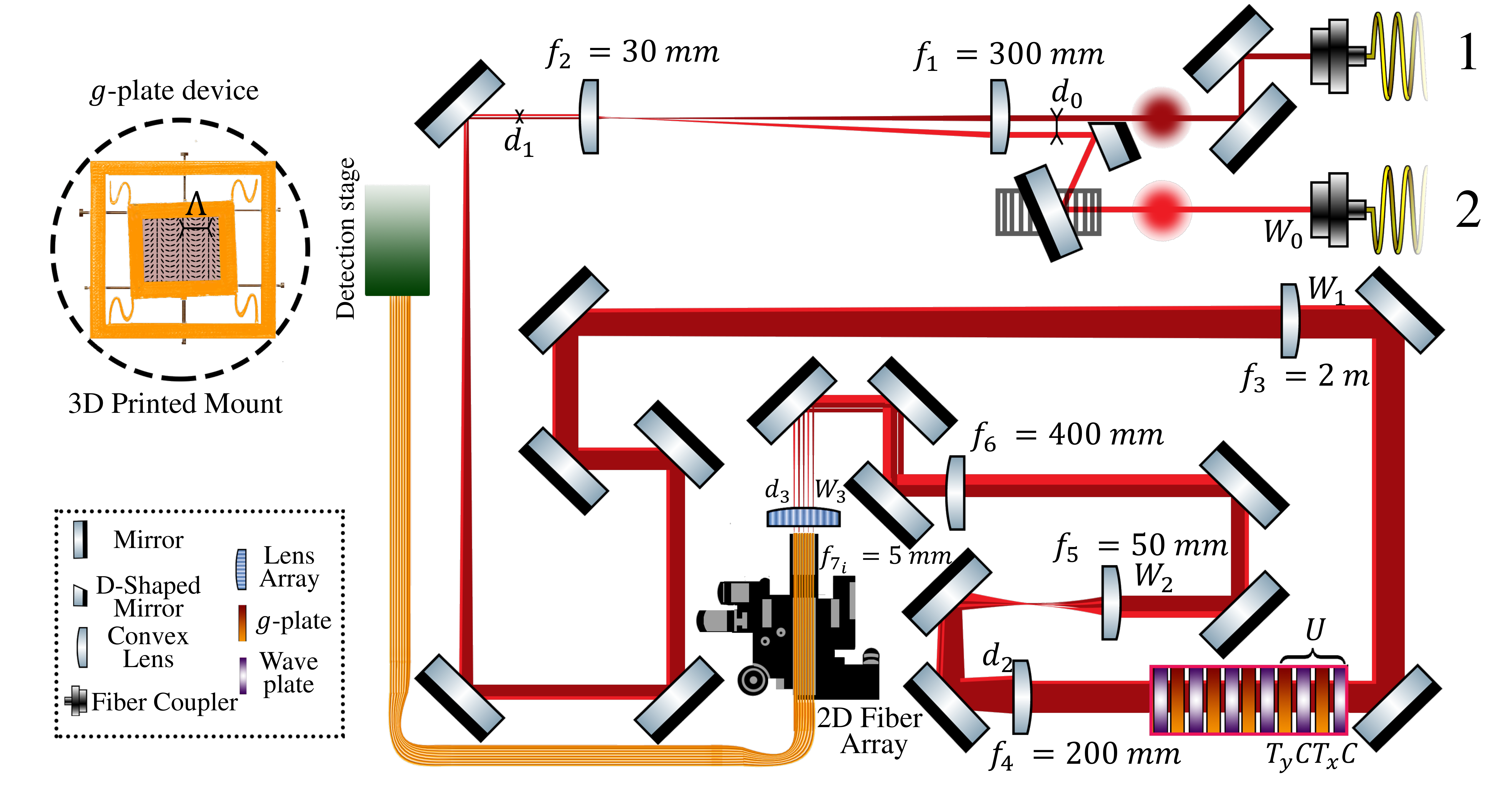}
   \caption{\textbf{Scheme of the 2 dimensional quantum walk apparatus}. In the single-particle regime, one photon was sent to the fiber coupler ``1". For 2D-QW with two walkers, the photons entered the system through ports ``1" and ``2". A system composed of several mirrors and lenses was properly designed to ensure an efficient coupling into the fiber array, considering the physical constraints imposed by its pitch. After the photons propagate through the g-plates, a micro-lens array was placed immediately before the multimode fiber-array to increase the coupling efficiency.} 
   \label{fig:sketch}
\end{figure}

\noindent The two-dimensional quantum walk was performed by a cascade of g-plates oriented along the $x$ and $y$ directions, interspersed with waveplates. For $n$ g-plates$_{x}$ and $n$ g-plates$_{y}$, the maximum angular difference between two beams is $\theta_{max} = 2\,n\,\Delta\,\theta$. At the output of the quantum walk structure, we used a three lens system to decrease the relative distance between consecutive beams and reduce their waist. The $f_{4} = 200$ mm lens was conjugated with the $f_{5}= 50$ mm one to reduce the waist by a factor 4. Then, we used a $f_{6} = 400$ mm lens in order to set the subsequent relative distances to $d_{3}$ and each waist to $W_{3}$.
 
\noindent Given two adjacent beams at the output of the g-plate-system, the angular difference $\Delta\,\theta$ induces a spatial shift at the focal point of $f_{6}$:
\begin{equation}
\begin{split}
    d_{3} &= -\Delta\theta\frac{f_4 f_6}{f_5}, \\\
    W_{3} &\simeq \frac{\lambda}{\pi W_{2}}\,\frac{f_4f_6}{f_5}.
\end{split}
\end{equation}

\noindent Numerically, we obtained: $d_3 = 250$ $\mathrm{\mu}$m and $W_3 = 80$ $\mathrm{\mu}$m. Hence, the maximum angular shift became a spatial shift equal to $n \times 250$ $\mathrm{\mu}$m. Finally, we used a micro-lens array (see Supplementary Note \ref{microlenses}) to improve the coupling efficiency into the fiber array. We employed a fiber array defining a $8\times8$ grid, in which the fibers were separated by $250$ $\mathrm{\mu}$m. Each one of these multi-mode fibers has a core of $62.5$ $\mathrm{\mu}$m and its output was coupled to a single-photon avalanche detector.

\noindent The measured rate at the output of the SPDC source was approximately $60$ kHz single-photon count at each arm, and $6$ kHz two-fold events by using a coincidence window of $8$ ns. The overall recorded two-fold events after the first, second and third steps of the quantum walk were $46926$, $65611$ and $90608$. The corresponding coincidence-rates after 1, 2 and 3 steps were  {$56$} Hz, $11$ Hz and $7$ Hz, respectively. The estimated losses for of the optical apparatus are due to the following optical elements:
         {\begin{itemize}
        \item tunable plate ($15\%$ for each plate); 
        \item fiber-array coupling ($40\%$);
        \item delay lines ($40\%$ for each photon);
        \item bunching probabilities detection (around $50\%$) (See the Supplementary Note 7);
    \end{itemize}}
\noindent As such, the overall transmission efficiency of the QW evolution (that is considering all liquid crystal devices) is $0.85^N$, where $N$ is the number of plates. This yields an efficiency equal to $\simeq 14\%$ after three steps, where $N=12$. However, by putting a standard anti-reflection coating on plate surfaces, the transmission of a single cell can be increased at least to $95 \%$ and the overall efficiency to $\simeq 54\%$. 

\section{Experimental results for coherent light inputs}
 \noindent For the coherent light data acquisition, we inserted a beamsplitter between the last lens and the micro-lense array, and we positioned the CCD on the reflected path for the image acquisition. We show the raw pictures of one-photon distributions with coherent light in Figure (4b) of the main text.  The dark counts of the images are removed by averaging the intensity values of the pixels in a $10\times10$ pixel square that is far more then 15 lattice steps away from the visible spots. We find the center of each spot of the image by looking for the local maxima of each picture. Then, the waist of each spot is estimated by performing a two-dimensional Gaussian fit on the image data. We calculated the intensity of each light spot in the pictures by summing the intensity value associated to each pixel that belongs to the spot. Then, we divided the latter by the total value of the intensity. In this way, we reconstructed the probability distributions for each step.

\section{Photon synchronization through Hong-Ou-Mandel interference}
\label{HOM}
\noindent Temporal synchronization between the two-photon paths was performed via a Hong-Ou-Mandel \cite{Hong87} test. Three liquid-crystal plates were exploited as a beamsplitter, following a procedure similar to that reported in Ref. \cite{DAmbrosio19} with \textit{q-plates}. In particular, we turned on the plates in the following order: a g-plate along $x$, a waveplate tuned as $\omega=\pi/4$ and, finally, a g-plate along $x$. Two identical photons were sent in two adjacent modes $(m+1,0)$ and $(m-1,0)$ of the first g-plates with polarization $\ket{R}$ and $\ket{L}$, respectively. We only considered the motion along the $x$ axis and we indicated the state of the photon in the second quantization formalism as  $a^{\dagger}_{i_x,P}\ket{0}$, where $i_x$ is the $x$ coordinate on the lattice and $P$ is the polarization state. The action of the three plates, written in the creation-annihilation operation formalism, reads:
\begin{equation}
\begin{split}    
a^{\dagger}_{m+1,R} \rightarrow &  \: \cos{\frac{\delta_1}{2}}\cos{\frac{\delta_2}{2}}a^{\dagger}_{m+1,V}+\frac{\imath}{\sqrt{2}}\cos{\frac{\delta_1}{2}}\sin{\frac{\delta_2}{2}}(a^{\dagger}_{m+2,R}-a^{\dagger}_{m,L})+ \\ + & \imath \sin{\frac{\delta_1}{2}}\cos{\frac{\delta_2}{2}}a^{\dagger}_{m,H}-\frac{1}{\sqrt{2}}\sin{\frac{\delta_1}{2}}\sin{\frac{\delta_2}{2}}(a^{\dagger}_{m+1,R}+a^{\dagger}_{m-1,L}),
\end{split}
\end{equation}
and:
\begin{equation}
\begin{split}
a^{\dagger}_{m-1,L} \rightarrow & \cos{\frac{\delta_1}{2}}\cos{\frac{\delta_2}{2}}a^{\dagger}_{m-1,H}+\frac{\imath}{\sqrt{2}}\cos{\frac{\delta_1}{2}}\sin{\frac{\delta_2}{2}}(a^{\dagger}_{m,R}+a^{\dagger}_{m-2,L})+ \\ + & \imath\sin{\frac{\delta_1}{2}}\cos{\frac{\delta_2}{2}}a^{\dagger}_{m,V}-\frac{1}{\sqrt{2}}\sin{\frac{\delta_1}{2}}\sin{\frac{\delta_2}{2}}(a^{\dagger}_{m+1,R}-a^{\dagger}_{m-1,L}). 
\end{split}
\end{equation}
Hence, the two-photon probability of having a bunching event in the initial positions $x_1=m+1$ and $x_2=m-1$ becomes:
\begin{equation}
P_{b}(m+1,m-1)=P(m+1,m+1)+P(m-1,m-1)=\left(\cos\frac{\delta_1}{2}\cos\frac{\delta_2}{2}\sin\frac{\delta_1}{2}\sin\frac{\delta_2}{2}\right)^2+\left(\sin{\frac{\delta_1}{2}}\sin{\frac{\delta_2}{2}}\right)^2.
\end{equation}

\noindent In Supplementary Figure \ref{fig:HOM} (a) we show the plots of the bunching probability as a function of the tuning parameters $\delta_1$ and $\delta_2$. For $\delta_1=\delta_2=\pi$ the probability of bunching is equal to $1$. For this value of the g-plates tuning, the plates mimic a beamsplitter with transitivity $t=1/\sqrt{2}$, in the subspace of g-plate modes $(m+1,R), (m-1,L)$:
\begin{equation}
M_G=\begin{pmatrix}
1/\sqrt{2} & 1/\sqrt{2}\\
1/\sqrt{2}& -1/\sqrt{2}
\end{pmatrix},
\end{equation}
that is the matrix representation of a 50/50 beamsplitter.

\noindent We experimentally performed the Hong-Ou-Mandel experiment in this configuration. The two photon input state was prepared in lattice sites $(-1,0)$ and $(1,0)$ with $\ket{L}$ and $\ket{R}$ polarization, respectively. In particular, we considered one of the two possible outputs of this unitary evolution to measure the bunching configuration. Indeed, for any arbitrary unitary operation, the probability of bunching of two indistinguishable photons exiting from the same output is twice the one obtained with distinguishable ones. More specifically, by considering a generic unitary $U$ and two photons entering in input modes $n$ an $m$, the probability that the two particles exit from the same output $k$ in the distinguishable and indistinguishable cases are respectively:
\begin{eqnarray}
    P_\text{dis}(k,k)=|U_{n,k}|^2|U_{m,k}|^2\\
    P_\text{ind}(k,k)= 2 |U_{n,k}|^2|U_{m,k}|^2,
\end{eqnarray}
independently of the unitary matrix. For this reason, the visibility of the HOM peak is independent of the g-plate unitary transformation and only depends on photon indistinguishability. To discriminate the events when two photons exit from the same output we used a 50/50 fiber beamsplitter (FBS). Then, we connected the FBS output ports to two avalanche photo-diodes (APDs). We then acquired the coincidence-rate at different positions of the translation stage. The measured interference pattern is shown in Supplementary Figure \ref{fig:HOM} (b). The visibility $v$ of the peak was estimated by using the following formula:
\begin{equation}
   v = \frac{cc_{\text{in}}-cc_{\text{out}}}{cc_{\text{out}}},
\end{equation}
where the $cc_{\text{in}}$ and $cc_{\text{out}}$ are the coincidence-rates at the maximum and large time separation, respectively. The obtained value of the visibility is $0.95\pm 0.02$, showing that the degree of indistinguishability of the photons is not significantly affected by the cascade of g-plates corresponding to the quantum walk platform.

\begin{figure}[t!]
    \includegraphics[width=\textwidth]{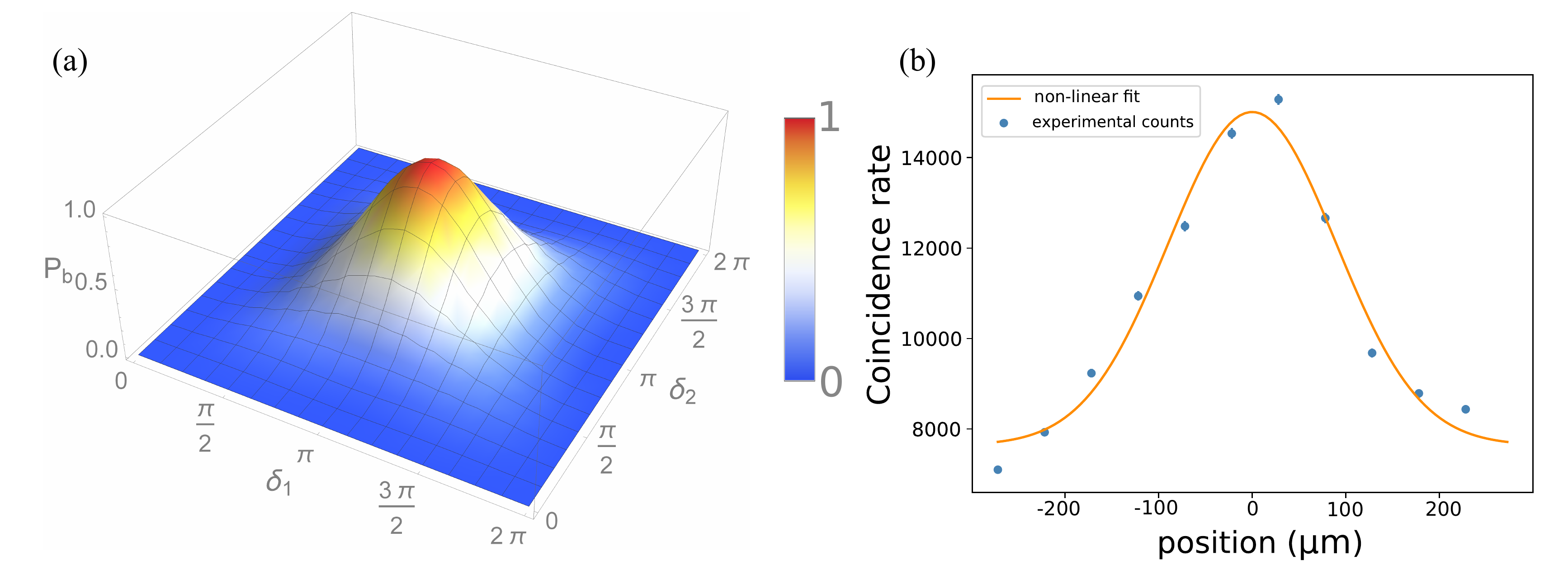}
   \caption{\textbf{Hong-Ou-Mandel test for photon path synchronization}. (a) We plot the probability of bunching ($P_b$) in position $(-1,0)$ and $(1,0)$ as a function of the electrical tuning $\delta_1$ and $\delta_2$ of the g-plates. We observe that the bunching probability approaches 1 when $\delta_1=\delta_2=\pi$. (b) Experimental verification of photon indistinguishability through Hong-Ou-Mandel (HOM) interference after the quantum walk platform. The poissonian statistical uncertainties are included in the experimental points. The measured visibility of the pattern is $v = 0.95\pm0.02$.}
   \label{fig:HOM}
\end{figure}

\section{Experimental results for one- and two- photon inputs}
\label{section:ExpResults}
\noindent Three steps of the $U$ 2D-QW protocol were performed with one and two photons. {The one-particle probability distributions are measured by counting the two-fold events between the photon that performs the quantum walk and the trigger photon. Conversely, the two-photon distributions are obtained from the two-fold events between each pair of output modes. We considered only the modes whose coincidence number is much higher than accindental ones (2  {modes} for the first step, 6 for the second step and 12 for the third step), as analogously performed in previous experiments \cite{Giordani2019,Crespi2013,Pitsios2017}. In order to measure also the bunching probability we use a fiber beamsplitter for each measured mode. Hence we use two avalanche photodiode detectors for each mode. The fiber beamsplitter allows one to detect bunching configurations, i.e. two photon per mode, with a probability $p=2t(1-t)$, where $t$ is the fiber beamsplitter transmissivity.  We calculate the error on each distribution value by using a bootstrapping approach. Then each probability distribution $P$ is obtained by dividing the two-fold events of each pair of modes by the total number of measured events. We remove the accidental coincidences (due to the dark counts) by considering the dark signal as a continuous stochastic signal. The accidental coincidences for each pair of modes $A_{i,j}$ are given by \cite{Janossy1944}:
  \begin{equation*}
    A_{i,j}=2S_iS_jTw
   \end{equation*}
   where $S_i$ and $S_j$ are the values of single-event rates of the $i$ and $j$ modes, $T$ is the total time of the data acquisition and $w$ is the coincidence windows, i.e. the time interval in which we consider two events as simultaneous. Finally, we correct each value with the coupling efficiencies. These corrections are obtained by injecting coherent laser light into each output mode and by calculating the coupling efficiency as the ratio between the output and the input light intensity.}

\noindent Here, we report the details on the similarity calculation for one- and two-particle regimes. Moreover, we report also the distributions for distinguishable particles. Finally, we show the plots of the non-classicality test, presented in Supplementary Equation \eqref{eqn:viol}, in the indistinguishable regime for all the steps. The errors on distributions values and all the derived quantities were calculated via bootstrapping approach, by considering that the distributions values are sampled on a Poissonian probability distribution.

\subsection{One-particle regime} 
\noindent In the one-particle regime, the experiment was carried out by directly detecting one of the two-photons, acting as a trigger, while the other is injected in the quantum-walk platform.
The probability distributions were measured at the end of each step, and the steps were performed by inserting the photon with polarization $\ket{D}$ in $(1,0)$. The distributions are shown in the main text. We compared the theoretical and experimental distributions using the similarity defined as:
\begin{equation}
    \mathcal{S}_{1\mathrm{p}}^{(t)}=\left(\sum_{\mathbf{r}}\sqrt{P^{(t)}(\mathbf{r})\tilde{P}^{(t)}(\mathbf{r})}\right)^2,
        \label{eqn:sim}
\end{equation}
where $t$ is the step number, $\mathbf{r}$ is the particle position on the lattice and $P^{(t)}(\mathbf{r})$ and $\tilde{P}^{(t)}(\mathbf{r})$ are the theoretical and the experimental distributions, respectively.

\subsection{Two-particle regime} 
\begin{figure}[t!]
\includegraphics[width=\textwidth]{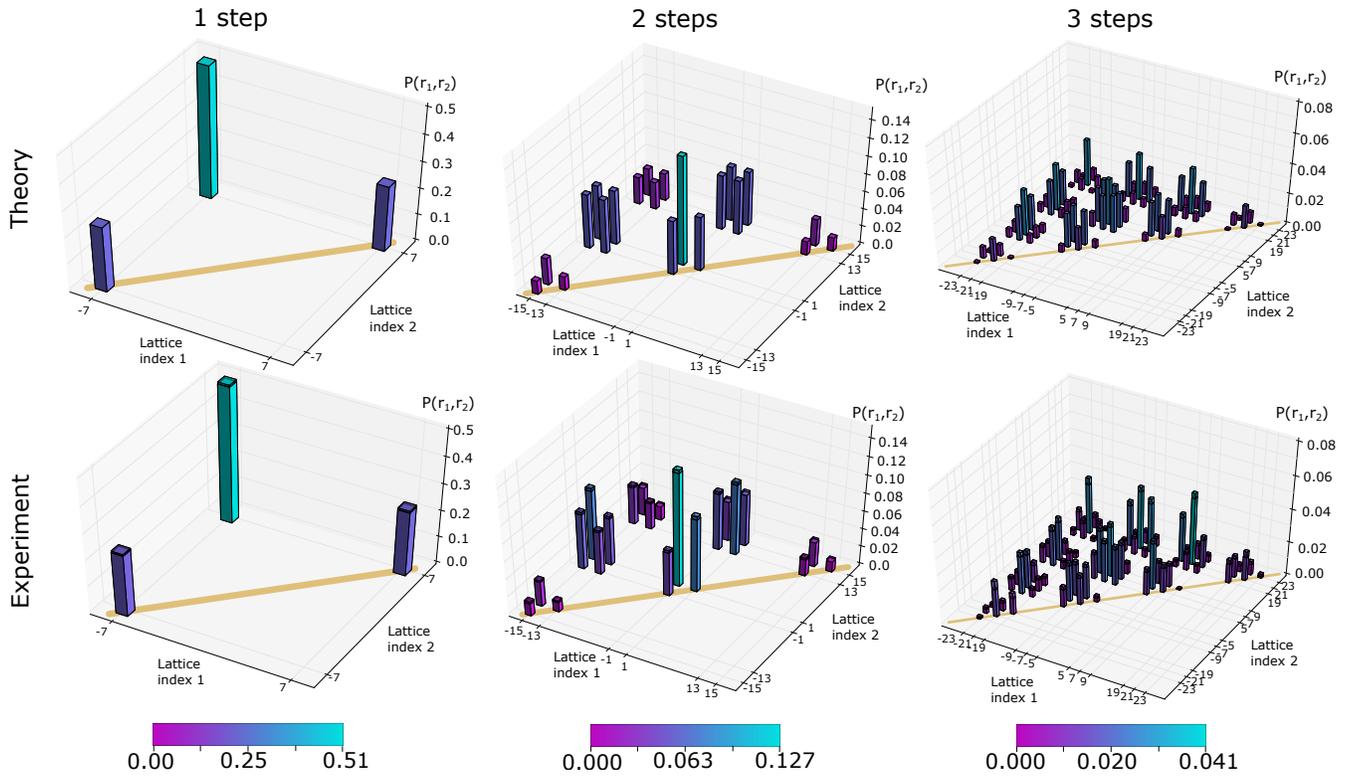}
\caption{\textbf{Theoretical distribution and experimental reconstruction of the 2D-QW with distinguishable particles.} The theoretical distribution was computed by setting $c_0=0$, corresponding to the visibility of distinguishable particles. The same linearization procedure employed in the main text for indistinguishable particles was performed to obtain a representation of the experimental data. Shaded regions on top of each bar correspond to the experimental error at one standard deviation.  The error bars were obtained thorough a bootstrapping approach. The bunching probabilities are highlighted by the yellow line.}
    \label{fig:sim2Pdist}
\end{figure}

\noindent In the two-particle regime, the photon paths were synchronized in the g-plate by performing a Hong-Ou-Mandel test (see Supplementary Figure \ref{fig:HOM}). As explained in Supplementary Note \ref{HOM}, such test was performed by exploiting the g-plates as a beamsplitter. The value of the measured visibility was employed to set the value of $c_0=0.95$ in the distribution theoretical calculations. Similarly to the one-particle regime, the probability distributions were measured at the end of each step. The steps were performed by injecting the two photons with polarization $\ket{A}$ and $\ket{D}$ in $(-1,0)$ and $(1,0)$, respectively. The two-dimensional lattice was linearized to perform a representation of the four-dimensional probabilities, thus presenting the latter quantities as two-parameter functions. The linearization of each lattice site was obtained as the following map $\mathbf{r} \rightarrow m + 7n$ with $m,n \in [-3,3]$ . To report the experimental coincidence-rate matrix, the same linearization was performed.
In the main text, we show distributions after each of the three steps for indistinguishable particles, with the corresponding value of similarity. The similarity was calculated by generalizing that corresponding to one particle in the following way:
\begin{equation}
    \mathcal{S}_{2\mathrm{p}}^{(t)}=\left(\sum_{\mathbf{r}_1,\mathbf{r}_2}\sqrt{P^{(t)}(\mathbf{r}_1,\mathbf{r}_2)\tilde{P}^{(t)}(\mathbf{r}_1,\mathbf{r}_2)}\right)^2,
        \label{eqn:sim2}
\end{equation}
where $t$ is the step number, $\mathbf{r}_1$ and $\mathbf{r}_2$ are the positions on the lattice of the particle 1 and 2 and $P^{(t)}(\mathbf{r}_1,\mathbf{r}_2)$ and $\tilde{P}^{(t)}(\mathbf{r}_1,\mathbf{r}_2)$ are the theoretical and the experimental distributions, respectively. 

\noindent In Supplementary Figure \ref{fig:sim2Pdist}, the theoretical and experimental distributions for distinguishable particle ($c_0=0$) for $1$, $2$ and $3$ steps are shown. The calculated similarities are $\mathcal{S}_{\mathrm{dis}}^{(1)}=0.9997\pm0.0001$, $\mathcal{S}_{\mathrm{dis}}^{(2)}=0.9886\pm0.0006$ and $\mathcal{S}_{\mathrm{dis}}^{(3)}=0.954\pm0.002$, thus proving a good agreement between the expected results and experimental ones.

\subsection{Violation Inequality} 
\noindent In order to estimate the \textit{non-classicality} of our experimental distribution, we employed the inequality introduced in Supplementary Ref. \cite{Bromberg2009,Peruzzo10}. More specifically, classical light has to satisfy the following condition:
\begin{equation}
    \mathcal{V}({\mathbf{m}_{1},\mathbf{m}_{2}}) =  {\frac{2}{3}} \sqrt{\Gamma_{\mathbf{m}_{1},\mathbf{m}_{1}}^{(\rm{cl})}\Gamma_{\mathbf{m}_{2},\mathbf{m}_{2}}^{(\rm{cl})}}-\Gamma_{\mathbf{m}_{1},\mathbf{m}_{2}}^{(\rm{cl})} < 0.
    \label{eqn:viol}
\end{equation}
where $\Gamma_{\mathbf{m}_1,\mathbf{m}_2}^{(\rm{cl})}$ is the classical probability that the light exits from the $\mathbf{m}_1$ and $\mathbf{m}_2$ output ports of an interferometer. The action of the interferometer is described by a unitary matrix $\mathcal{U}$. The evolution of 2D-QW with g-plates is equivalent to the action of an interferometer with unitary matrix given by $\mathcal{U}=U^t$ at each step $t$. The output modes of 2D-QW interferometer are equivalently given by $\mathbf{m}={\mathbf{r},\sigma}$ where $\mathbf{r}$ is the position on the lattice and $\sigma= {\uparrow,\downarrow}$. By measuring the final distribution of 2D-QW we trace away the polarization state. For this reason we do not have a direct access to $\Gamma_{\mathbf{r}_1,\sigma_1,\mathbf{r}_2,\sigma_2}$. Hence, we recast Supplementary Equation \eqref{eqn:viol} in terms of the measured values of $\Gamma_{\mathbf{r}_1,\mathbf{r}_2} = \sum_{\sigma_1,\sigma_2}\Gamma_{\mathbf{r}_1,\sigma_1,\mathbf{r}_2,\sigma_2}$.
In order to do that, we sum the violation inequalities for the fixed positions $\mathbf{r}_1$ and $\mathbf{r}_2$ on the lattice:

\begin{equation}
   \sum_{\sigma_1,\sigma_2}\mathcal{V}({\mathbf{r}_1,\sigma_1,\mathbf{r}_2,\sigma_2})=  {\frac{2}{3}} \sum_{\sigma_1,\sigma_2}\sqrt{\Gamma_{\mathbf{r}_1,\sigma_1,\mathbf{r}_1,\sigma_1}^{(\rm{cl})}\Gamma_{\mathbf{r}_2,\sigma_2,\mathbf{r}_2,\sigma_2}^{(\rm{cl})}}-\Gamma_{\mathbf{r}_1,\mathbf{r}_2}^{(\rm{cl})} < 0.
    \label{eqn:violint}
\end{equation}

\noindent By exploiting the inequality:
\begin{equation}
    \sqrt{\sum_i{a}_i}<\sum_i\sqrt{a_i},            
\end{equation}
we can write the Supplementary Equation \eqref{eqn:violint} as:
\begin{equation}
     {\frac{2}{3}}\sqrt{ \sum_{\sigma_1}\Gamma_{\mathbf{r}_1,\sigma_1,\mathbf{r}_1,\sigma_1}^{(\rm{cl})}\sum_{\sigma_2}\Gamma_{\mathbf{r}_2,\sigma_2,\mathbf{r}_2,\sigma_2}^{(\rm{cl})}}-\Gamma_{\mathbf{r}_1,\mathbf{r}_2}^{(\rm{cl})} < 0.
    \label{eqn:violint2}
\end{equation}

\noindent By using the definition of $\Gamma_{\mathbf{r}_1,\mathbf{r}_1}$ and $\Gamma_{\mathbf{r}_2,\mathbf{r}_2}$ it can be further rewritten as:
\begin{equation}
     {\frac{2}{3}}\sqrt{ (\Gamma_{\mathbf{r}_1,\mathbf{r}_1}^{(\rm{cl})}-\Gamma_{\mathbf{r}_1,\uparrow,\mathbf{r}_1,\downarrow}^{(\rm{cl})}-\Gamma_{\mathbf{r}_1,\downarrow,\mathbf{r}_1,\uparrow}^{(\rm{cl})})(\Gamma_{\mathbf{r}_2,\mathbf{r}_2}^{(\rm{cl})}-\Gamma_{\mathbf{r}_2,\uparrow,\mathbf{r}_2,\downarrow}^{(\rm{cl})}-\Gamma_{\mathbf{r}_2,\downarrow,\mathbf{r}_2,\uparrow}^{(\rm{cl})})}-\Gamma_{\mathbf{r}_1,\mathbf{r}_2}^{(\rm{cl})} < 0.
    \label{eqn:violint3}
\end{equation}

\noindent Finally, considering that $\Gamma_{\mathbf{r},\uparrow,\mathbf{r},\downarrow} = \Gamma_{\mathbf{r}\downarrow,\mathbf{r},\uparrow}$ we can write:

\begin{equation}
    \mathcal{V}({\mathbf{r}_1,\mathbf{r}_2})= {\frac{2}{3}}\sqrt{ \Gamma_{\mathbf{r}_1,\mathbf{r}_1}^{(\rm{cl})}\Gamma_{\mathbf{r}_2,\mathbf{r}_2}^{(\rm{cl})}-2\Gamma_{\mathbf{r}_2,\mathbf{r}_2}^{(\rm{cl})}\Gamma_{\mathbf{r}_1,\uparrow,\mathbf{r}_1,\downarrow}^{(\rm{cl})}-2\Gamma_{\mathbf{r}_1,\mathbf{r}_1}^{(\rm{cl})}\Gamma_{\mathbf{r}_2,\uparrow,\mathbf{r}_2,\downarrow}^{(\rm{cl})}}-\Gamma_{\mathbf{r}_1,\mathbf{r}_2}^{(\rm{cl})} < 0.
    \label{eqn:lastviol}
\end{equation}

\noindent In the case in which $\Gamma_{\mathbf{r}_1,\uparrow,\mathbf{r}_1,\downarrow}$ and $\Gamma_{\mathbf{r}_2,\uparrow,\mathbf{r}_2,\downarrow}$ are theoretically zero, we can use this violation to demonstrate the quantumness of the output  distribution. 

\noindent We applied this non-classicality test to the measured output distributions for indistinguishable particle inputs, shown in the main text. More specifically, we calculated the value $\mathcal{V}(\mathbf{r}_{1},\mathbf{r}_{2})$, using Supplementary Equation \eqref{eqn:lastviol}, and the corresponding error $\sigma_{V}(\mathbf{r}_{1},\mathbf{r}_{2})$, by propagating the experimental error in the distribution.  In Figure 6 of the main text we report the complete plots of $\mathcal{V}(\mathbf{r}_{1},\mathbf{r}_{2})/\sigma_{V}(\mathbf{r}_{1},\mathbf{r}_{2})$  {for the second and the third step, respectively}. It is worth to notice that we compute the violation only for the case in which the relation $\Gamma_{\mathbf{r}_1,L,\mathbf{r}_1,R}=\Gamma_{\mathbf{r}_2,L,\mathbf{r}_2,R}=0$ is satisfied. The presence of output configurations that violate inequality \eqref{eqn:lastviol} at each step testifies the non-classicality of the obtained distributions. In Table I of the main text, the value of $\mathcal{V}(\mathbf{r}_{1},\mathbf{r}_{2})/\sigma_{V}(\mathbf{r}_{1},\mathbf{r}_{2})$ corresponding to the configuration that provides maximum violation is reported.\\